\documentclass[12pt,reqno]{amsart}
\usepackage{amsmath,amssymb,amsfonts,amsthm}
\usepackage[mathscr]{eucal}
\usepackage[all]{xy}
\usepackage{hyperref}
\usepackage{setspace}
\allowdisplaybreaks

\author{V.A.~Abakumova, I.Yu.~Karataeva, S.L.~Lyakhovich}

\address{Physics Faculty, Tomsk State University, Lenin ave. 36, Tomsk 634050, Russia.}

\allowdisplaybreaks

\title{Unfree gauge symmetry in the Hamiltonian formalism }

\begin{document}

\maketitle

\begin{abstract}
The constrained Hamiltonian formalism is worked out for the theories
where the gauge symmetry parameters are unfree, being restricted by
differential equations. The Hamiltonian BFV-BRST embedding is
elaborated for this class of gauge theories. The general formalism
is exemplified by the linearized unimodular gravity.
\end{abstract}

\section{Introduction}
\noindent
If the gauge variation of action identically vanishes under the
condition that the gauge parameters obey differential equations, the
gauge symmetry is said unfree. The general structure of  unfree
gauge symmetry algebra has been recently established in ref
\cite{KAPARULIN2019114735}. The extension of the BV (Batalin-Vilkovisky)
formalism to  unfree gauge symmetry is proposed in
\cite{Kaparulin2019}.

The most known example of unfree gauge symmetry is provided by
unimodular gravity where the diffeomorphism parameters
$\epsilon^\mu$ are constrained by transversality condition
$\nabla_\mu\epsilon^\mu=0$. For discussion of consequences of
transversality condition in the unimodular gravity and various
extensions, see \cite{BUCHMULLER1988292}, \cite{BUCHMULLER1989313}, \cite{PhysRevD.40.1048},
\cite{Ellis_2011}, \cite{Gielen_2018}, \cite{PhysRevD.97.026007},
\cite{Percacci2018}, \cite{BARVINSKY201759}, \cite{PhysRevD.100.023542} and
references therein. More examples of  unfree gauge symmetry can be
found among multifarious higher spin field theories, see
\cite{SKVORTSOV2008301}, \cite{Campoleoni2013},
\cite{FRANCIA2014248}, \cite{Francia2017}.

While the phenomenon of unfree gauge symmetry is
well-known in terms of Lagrangian formalism,  it has been so far
unclear how the unfree symmetry reveals in the corresponding
Hamiltonian formalism. Even in well-studied models, like
unimodular gravity, the transversality condition on the
diffeomorphism parameters is not evident from the viewpoint of the
Poisson algebra of Hamiltonian constraints. This problem is noticed
in the literature, see \cite{Gielen_2018} and references therein.

In this article, we work out the general Hamiltonian description of
unfree gauge symmetry. In section 2, we list the basic features
of Lagrangian description of general unfree gauge algebra as
this is essential for constructing the Hamiltonian analogue. In 
section 3, we establish the general structure of unfree gauge
symmetry in the constrained Hamiltonian formalism; in section 4,
we construct corresponding Hamiltonian BRST
(Becchi-Rouet-Stora-Tyutin) complex; the section 5 exemplifies the
general formalism by the model of linearized unimodular gravity.

\section{Generalities of unfree gauge symmetry in Lagrangian formalism}

\noindent
In the reference \cite{KAPARULIN2019114735}, it is noticed that the
unfree gauge algebra is generated by four key ingredients: the
action functional $S$; the generators of unfree gauge symmetry
$\Gamma^i_\alpha$; the mass-shell completion functions $\tau_a$; the
operators of gauge parameter constraints $\Gamma^a_\alpha$. The
first two generating structures -- the action, and the gauge
symmetry generators -- are the key ingredients of any gauge symmetry
algebra, be the gauge parameters constrained, or not. The other two
generating elements, $\tau_a$ and $\Gamma^a_\alpha$, are special for
the unfree gauge symmetry. Let us non-rigorously  explain their role
in the dynamics, for a more systematic exposition we refer to
\cite{KAPARULIN2019114735}, \cite{Kaparulin2019}.

The main distinctive feature of theories with the unfree gauge
symmetry is that the local quantities exist such that vanish on-shell
 while they are \emph{not} spanned by the l.h.s. of Lagrangian
equations (EoM's). In other words, the generating set for the ideal
of on-shell vanishing local functions is not exhausted by the l.h.s.
of EoM's, it includes some other quantities, denoted as $\tau_a$:
\begin{equation}\label{I}
T(\phi)\approx 0 \quad\Leftrightarrow\quad
T(\phi)=\theta^i(\phi)\,\partial_i S(\phi) \, + \,
\theta^a(\phi)\,\tau_a(\phi)\, ,
\end{equation}
where $\approx$ means on-shell equality, and $\theta (\phi)$ are
local. Here we use the DeWitt condensed notation. The local
quantities $\tau_a(\phi)$ are supposed independent, and they do not
reduce to a linear combination of Lagrangian equations,
$\tau_a\neq\theta_a^i\partial_iS$. The quantities $\tau_a$ are
called the mass-shell \emph{completion functions}. Examples of the
completion functions can be found in \cite{KAPARULIN2019114735}, 
\cite{Kaparulin2019}. In general,  modified Noether identities
involve both  Lagrangian equations and completion functions:
\begin{equation}\label{GI}
    \Gamma^i_\alpha(\phi)\,\partial_i S(\phi)
    \, +\,
    \Gamma^a_\alpha(\phi)\,\tau_a(\phi)\equiv 0\,.
\end{equation}
With appropriate regularity assumptions (see in
\cite{KAPARULIN2019114735}, \cite{Kaparulin2019}), relations
(\ref{I}), (\ref{GI}) define the unfree gauge symmetry of the
theory. In particular, the gauge variation of the fields,
\begin{equation}\label{GVar}
\delta_\epsilon\phi^i=\Gamma^i_\alpha (\phi)\,\epsilon^\alpha\,,
\end{equation}
is a symmetry of the action provided that the gauge parameters
$\epsilon$ obey the equations
\begin{equation}\label{GPC}
\Gamma^a_\alpha (\phi)\,\epsilon^\alpha=0\,.
\end{equation}
The operators of gauge parameter constraints $\Gamma^a_\alpha
(\phi)$ are supposed independent. This means that the kernel of
$\Gamma^a_\alpha (\phi)$ is, at maximum, finite dimensional,
\begin{equation}\label{KerG}
\Gamma^a_\alpha (\phi)u_a=0 \quad \Rightarrow\quad u_a\in
M=\textmd{Ker}\,\Gamma^a,\quad \dim{M}=n\in\mathbb{N} \, .
\end{equation}
Here, $M$ is understood as a moduli space of the field theory. Given
(\ref{GI}), (\ref{KerG}), on shell $\tau_a\approx\Lambda_a,
\Lambda_a\in M$. In principle, the modular parameters $ \Lambda_a$
can be included into the definition of $\tau_a$, so the completion
functions can be considered on-shell vanishing without loss of
generality. Also notice that the relations $\tau_a\approx 0$ hold
true for any solution of the EoM's with corresponding modular
parameter, though these relations are not differential consequences
of EoM's.

The modified Noether identities (\ref{GI}) along with corresponding
regularity assumptions lead to the compatibility conditions
involving higher structures, which define the full unfree gauge
symmetry algebra. This algebra is more general than the one with
unconstrained gauge parameters. Corresponding gauge formalism is
worked out in references \cite{KAPARULIN2019114735},
\cite{Kaparulin2019}, we do not address it here, listing only the
basic relations.

\textbf{Example}. Let us illustrate relations
(\ref{GI}), (\ref{GPC}), (\ref{KerG}) in uncondensed notation by an
example. Consider the theory of fields
$\phi^{\texttt{i}}(x)$, where $\texttt{i}$ is a
discrete index (it can be spinorial, tensorial, isotopic index), and
$x^\mu$ are coordinates of space-time. Suppose the differential
consequences of EoM's can be linearly combined into the gradient of
 local quantity $\tau$, being the scalar function of fields and
their first derivatives with respect to $x^\mu$:
\begin{equation}\label{GI-eg}
    \hat{\Gamma}{}^{\texttt{i}}_\mu(\phi,\partial\phi)\, \frac{\delta S[\phi]}{\delta
    \phi^{\texttt{i}}}\,+\, \partial_\mu\tau (\phi,\partial\phi)\, \equiv\, 0 \,
    ,\quad \tau (0,0)\equiv 0\,,
\end{equation}
where $\hat{\Gamma}{}^{\texttt{i}}_\mu$ is a matrix whose
entries are linear differential operators with the field-depending
coefficients. As the derivatives of $\tau$ vanish on-shell,  it is
an on-shell constant, $\tau(\phi,\partial\phi)\approx\Lambda=const$.
The specific value of the constant $\Lambda$ is determined by the
boundary conditions or asymptotics of fields, not by initial data.
For example, if the fields are supposed vanishing at spacial
infinity of the space-time, then $\Lambda=0$ irrespectively to the
initial data, as $\tau(0,0)=0$. So $\Lambda$ should be understood as
a modular parameter rather than an integral of motion. This modular
parameter can be included into definition of the completion
function, so one can set $\tau\approx 0$ without restricting
generality. Once the modified Noether identity (\ref{GI}) reads as
(\ref{GI-eg}) in this example, the gauge parameter should be the
vector field $\epsilon^\mu(x)$. The gauge parameter constraint
operator $\Gamma^a_\alpha$ of (\ref{GI}) is identified by
(\ref{GI-eg}) as $\partial_\mu$, so the equation (\ref{GPC}) reduces
to the transversality condition $\partial_\mu\epsilon^\mu=0$ for the
gauge parameter $\epsilon^\mu$. The unfree gauge transformations
(\ref{GVar}) are generated by the conjugate to the operator
$\hat{\Gamma}$ involved in the modified Noether identity (\ref{GI}):
\begin{equation}\label{Gamma-dagger}
\delta_\epsilon\phi^{\texttt{i}}=\hat{\Gamma}{}^{\dagger\texttt{i}}_\mu\epsilon^\mu\,,
\quad \partial_\mu\epsilon^\mu\equiv 0 \,.
\end{equation}
Given the modified Noether identity (\ref{GI-eg}), the
transformation (\ref{Gamma-dagger}) leaves the action invariant:
\begin{equation}\label{GS-eq}
    \delta_\epsilon S\equiv\int dx\,\frac{\delta S[\phi]}{\delta \phi^{\texttt{i}}}
    \hat{\Gamma}{}^{\dagger\texttt{i}}_\mu(\phi,\partial\phi)\epsilon^\mu
    \equiv \int  dx \,\epsilon^\mu\hat{\Gamma}{}^{\texttt{i}}_\mu(\phi,\partial\phi)\frac{\delta S[\phi]}{\delta \phi^{\texttt{i}}}
    \equiv \int dx \,(\partial_\mu\epsilon^\mu)\tau(\phi,\partial\phi)\equiv
    0 \, .
\end{equation}
The unimodular gravity is covered by this example. The role of the
modular parameter is played by the cosmological constant, while the
completion function is the scalar curvature. Unfree gauge symmetries
of some higher spin field theories, see \cite{SKVORTSOV2008301},
\cite{Campoleoni2013}, \cite{FRANCIA2014248},
\cite{Francia2017}, also follow the pattern of this example,
though the completion functions are tensors in these models, not
scalars.

\section{Unfree gauge symmetry transformations in the Hamiltonian formalism}

\noindent
Consider the action of Hamiltonian
theory with primary constrains,
\begin{equation}\label{SH}
S=\int dt\left(p_i\dot{q}{}^i-H_T(q,p,\lambda)\right)\,, \quad
H_T(q,p,\lambda)= H(q,p)+\lambda^\alpha T_\alpha(q,p)\, .
\end{equation}
The role of fields here is played by canonical variables $q^i,p_i$,
and Lagrange multipliers $\lambda^\alpha$. In the previous section,
the structures are described that define the unfree gauge symmetry
for the general action. In this section, we detail these structures
for the specific action  (\ref{SH}) and find thereby the Hamiltonian
form of unfree gauge symmetry transformations.

For the action (\ref{SH}), the EoM's read
\begin{equation}
\label{HE}
 \frac{\delta S}{\delta p_i}\equiv\dot{q}{}^i -\{q^i,H_T\} =0\, , \quad \frac{\delta S}{\delta q^i}\equiv -\,\dot{p}_i +\{p_i,H_T\} =0\,;
\end{equation}
\begin{equation}\label{T}
\frac{\delta S}{\delta \lambda^\alpha}\equiv -\,T_\alpha(q,p)=0 \, .
\end{equation}
The constraints $T_\alpha(q,p)$ are supposed irreducible. At this
point we accept an auxiliary assumption that the differential
consequences of the equations do not fix $\lambda^\alpha$ as
functions of $q,p$. Assuming that the Dirac conjecture holds true
for theory (\ref{SH}), this means there are no second-class
constraints.

Now, our primary objective is to identify completion functions
(\ref{I}) and gauge identities (\ref{GI}) for EoM's (\ref{HE}),
(\ref{T}). We begin with applying the Dirac-Bergmann algorithm to
equations (\ref{HE}), (\ref{T}):
\begin{equation}\label{dotT}
    \dot{T}_\alpha \approx \{T_\alpha, H_T\} \approx 0 \, .
\end{equation}
Once the Lagrange multipliers are not defined by conservation of the
primary constraints, the r.h.s. of the above relation should be a
linear combination of the primary constraints and the secondary
ones. Let us assume that the irreducible generating set
$\{\tau_a(q,p)\}$ can be chosen for the secondary constraints. This
means the local differential operators $\Gamma, W$ exist such that
\begin{equation}\label{GVtau}
    \{T_\alpha(q,p)), H_T
    (q,p,\lambda))\}=W_\alpha^\beta(q,p,\lambda)\,T_\beta(q,p)\,+\,
    \Gamma_\alpha^a(q,p,\lambda)\,\tau_a(q,p)\, .
\end{equation}
The irreducibility assumption of  the secondary constraints $\tau_a$
is twofold. First, all the constraints should be independent:
\begin{equation}\label{irred}
    \theta^\alpha T_\alpha\,+\,
    \theta^a \tau_a=0\quad\Leftrightarrow\quad \theta^\alpha
    =A^{\alpha\beta}T_\beta\,+\,A^{\alpha a}\tau_a \,, \quad\theta^a=A^{ab}\tau_b\, -\,A^{\alpha
    a}T_\alpha\,,
\end{equation}
where $    A^{ab}=-A^{ba},\, A^{\alpha\beta}=-A^{\beta\alpha}
$ . Second, the kernel of the operator $\Gamma^a_\alpha$ should be,
at most, finite dimensional, in the sense of relation (\ref{KerG}).
The only difference with the Lagrangian formalism is that $\Gamma$ in
(\ref{GVtau}) involves derivatives only by space coordinates, while
the Lagrangian counterpart can differentiate also by time.

If $\Gamma^a_\alpha$ admitted the dual \emph{differential} operator
$\tilde\Gamma^\alpha_a$ such that
\begin{equation}\label{tildeG}
    \tilde{\Gamma}{}^\alpha_a\,\Gamma_\alpha^b=\delta^b_a \,,
\end{equation}
then the secondary constraints $\tau_a$ would be the differential
consequences of the original equations (\ref{HE}), (\ref{T}). In the
opposite case, $\tau_a$ reduce to the element of the kernel of the
differential operator $\Gamma$. In this case $\tau$ are considered
as completion functions, and hence the gauge symmetry should be
unfree. Once the kernel is finite of $\Gamma^a_\alpha$, completion
functions $\tau_a(q,p)$  can be redefined by adding modular
parameters $\Lambda_a$ to make $\tau$ vanishing on shell:
\begin{equation}\label{tau-Lambda}
    \Gamma^a_\alpha\,\tau_a=0\quad\Leftrightarrow\quad
    \tau_a=\Lambda_a, \quad \Lambda_a\in \textmd{Ker}\,
    \Gamma^a_\alpha\,; \quad \tau_a\,\mapsto \tau_a-\Lambda_a\,.
\end{equation}
Then $\tau_a$ still vanish on-shell and viewed as secondary
constraints, though they are \emph{not differential consequences} of
the EoM's, and hence the gauge symmetry should be unfree.

The next simplifying assumption is that no tertiary constraints
appear. This means that the time derivatives of secondary
constraints reduce on-shell to the combinations of themselves and
the primary ones:
\begin{equation}\label{dot-tau}
    \dot{\tau}_a\approx{}\{{\tau}_a,
    H_T\}=W^\alpha_a(q,p,\lambda)T_\alpha+\Gamma^b_a(q,p,\lambda)\tau_b.
\end{equation}
Off shell, the time derivatives of primary and secondary constraints
identically reduce to the linear combination of constraints and EoM's (\ref{HE}):
\begin{equation}\label{GIH-alpha}
\{T_\alpha , q^j\}\, \frac{\delta S}{\delta q^j} + \{T_\alpha ,
p_j\}\, \frac{\delta S}{\delta p_j}
+\Big(\delta^\beta_\alpha\frac{d}{dt}
-W_\alpha^\beta(q,p,\lambda)\Big)\,\frac{\delta S}{\delta
\lambda^\beta}+\Gamma_\alpha^a(q,p,\lambda)\,\tau_a \equiv 0\,;
\end{equation}
\begin{equation}\label{GIH-a}
\{\tau_a  , q^j\}\, \frac{\delta S}{\delta q^j} + \{\tau_a  ,
p_j\}\, \frac{\delta S}{\delta p_j}
-W^\alpha_a(q,p,\lambda)\frac{\delta S}{\delta \lambda^\alpha}+
\Big(- \delta^b_a \frac{d}{dt} + \Gamma^b_a(q,p,\lambda)\Big)\,
\tau_b \equiv 0 \, .
\end{equation}
Since the secondary constraints $\tau_a$ are not the
\emph{differential} consequences of the primary ones (\ref{T}), the
above relations are modified gauge identities (\ref{GI}) rather than
usual Noether identities between the variational equations. The
identities  (\ref{GI}) are equivalent to the unfree gauge symmetry
(\ref{GVar}) of the action, with the gauge parameters constrained by
the equations (\ref{GPC}). Given the gauge identities
(\ref{GIH-alpha}), (\ref{GIH-a}), the unfree gauge transformations
(\ref{GVar}) for constrained Hamiltonian system read
\begin{equation}
\label{GVar-H}
    \delta_\epsilon O(q,p)=\{O,T_\alpha \}\,\epsilon^\alpha\,+\, \{O,\tau_a\}\,\epsilon{\,}^a\,
    , \quad\delta_\epsilon\lambda^\alpha= \dot{\epsilon}{\,}^\alpha + W^\alpha_\beta(q,p,\lambda)\epsilon^\beta
    + W^\alpha_a(q,p,\lambda) \epsilon^a \,.
\end{equation}
The constraints on the gauge parameters (\ref{GPC}) are defined by
the coefficients at $\tau_a$ in the modified gauge identities
(\ref{GI}). Given specific identities (\ref{GIH-alpha}),
(\ref{GIH-a}), the constraints on gauge parameters read:
\begin{equation}\label{GPC-H}
\Big(\delta^b_a \frac{d}{dt} +
\Gamma^b_a(q,p,\lambda)\Big)\epsilon^a+\Gamma_\alpha^b(q,p,\lambda)\epsilon^\alpha=0
\, .
\end{equation}
The unfree gauge transformations (\ref{GVar-H}), (\ref{GPC-H}) have
been deduced above by using the gauge identities (\ref{GIH-alpha}),
(\ref{GIH-a}) for the theory (\ref{SH}) with the involution
relations (\ref{GVtau}), (\ref{dot-tau}).  By direct variation, one
can verify that the action (\ref{SH}) is indeed invariant under the
unfree gauge transformations (\ref{GVar-H}), (\ref{GPC-H}):
\begin{equation}\nonumber
\delta_\epsilon S \equiv  \int dt\, \Big( -\delta_\epsilon q^i
\,\big(\dot{p}_i -\{p_i,H_T\}\big) +\delta_\epsilon p_i \,
\big(\dot{q}{}^i -\{q^i,H_T\}\big) - \delta_\epsilon
\lambda^\alpha \, {T}_\alpha \Big)
\end{equation}
\begin{equation}\nonumber
\equiv\int dt\, \Big( - \dot{T}_\alpha\, \epsilon^\alpha -
\dot{\tau}_a\,\epsilon{\,}^a + \{ T_\alpha , H_{T}\}\,
\epsilon^\alpha +\{ \tau_a , H_{T}\}\,\epsilon{\,}^a - \big(
\dot{\epsilon}{\,}^\alpha +
W^\alpha_\beta(q,p,\lambda)\epsilon^\beta + W^\alpha_a(q,p,\lambda)
\epsilon^a\big) \, {T}_\alpha \Big)\, .
\end{equation}
Upon substitution $ \{ T_\alpha , H_{T}\}, \ \{ \tau_a , H_{T}\}\,$
from relations (\ref{GVtau}), (\ref{dot-tau}), the variation reads
\begin{equation}\label{direct-var}
\delta_\epsilon S \equiv \int dt \Big(\big(\dot{\epsilon}{\,}^a
+\Gamma_b^a(q,p,\lambda)\epsilon{\,}^b
+\Gamma_\alpha^a(q,p,\lambda)\epsilon^\alpha \big)\, \tau_a
-\frac{d}{dt}\big( T_\alpha \epsilon^\alpha + \tau_a \epsilon^a
\big) \Big)
 \, .
\end{equation}
Once the gauge parameters obey equations (\ref{GPC-H}), the
integrand reduces to the total derivative, so the action is indeed
invariant under the unfree gauge variation (\ref{GVar-H}),
(\ref{GPC-H}).

Let us discuss the constraints imposed on the gauge parameters
$\epsilon^\alpha$ and $\epsilon^a$ by equations (\ref{GPC-H}).
Equations (\ref{GPC-H}) define  $\dot{\epsilon}{}^a$ in terms of
$\epsilon^\alpha$. As the kernel of $\Gamma^a_\alpha$ is at maximum
finite in the sense of relation (\ref{KerG}), the time evolution of
$\epsilon^a$ is completely controlled by $\epsilon^\alpha$, while
the latter parameters are unconstrained by the equations. As the
equations (\ref{GPC-H}) have the structure
$\dot{\epsilon}{}^a=f^a(\epsilon^b,\epsilon^\beta)$, they admit any
initial data for $\epsilon^a$, so these parameters are arbitrary at
initial moment.

Alternatively, equations (\ref{GPC-H}) can be considered as
constraints imposed on the parameters $\epsilon^\alpha$, defining
some of them in terms of the rest of $\epsilon^\alpha$ and
$\dot{\epsilon}{}^a, \epsilon^a$. If all the constraints
(\ref{GPC-H}) are explicitly resolved by excluding some of the gauge
parameters $\epsilon^\alpha$, then the gauge transformations of
canonical variables (\ref{GVar-H}) will include
$\dot{\epsilon}{}^a$, while the variation of $\lambda^\alpha$ will
involve $\ddot{\epsilon}{}^a$. In this way, the first-order unfree
gauge symmetry (\ref{GVar-H}), (\ref{GPC-H}) is replaced by the second-order
 gauge symmetry with unconstrained gauge parameters. If the
spacial locality is not an issue, the constrained Hamiltonian
equations (\ref{HE}), (\ref{T}) always admit  the unconstrained
parametrization of gauge transformations with higher order
time derivatives of gauge parameters \cite{doi:10.1063/1.3193684}.  Also
notice that any linear system of local field equations admits
unconstrained local parametrization of gauge symmetry, possibly with
higher derivatives, though the transformations can be reducible
\cite{FRANCIA2014248}. So, all these facts lead to the conjecture
that the unfree gauge symmetry can be always equivalently replaced
by local higher order reducible gauge symmetry. This conjecture will
be addressed elsewhere.

Now, let us discuss the issue of on-shell gauge invariants. The long-known wisdom of Hamiltonian constrained dynamics about the gauge
invariants is that they should Poisson-commute on shell with all
first-class constraints, both primary and secondary \cite{dirac2001lectures}.
While unfree gauge transformations (\ref{GVar-H}), (\ref{GPC-H})
have been previously unknown, the gauge invariants turn out defined
in the same way as with unconstrained gauge parameters. This fact
can be seen from the above mentioned properties of equations
(\ref{GPC-H}). Let us explain that. Once any initial data for gauge
parameters $\epsilon^\alpha$, $\epsilon^a$ are admitted by the
equations (\ref{GPC-H}), the phase-space function $O(q,p)$ cannot be
invariant under the gauge transformation (\ref{GVar-H}) unless it
Poisson-commutes with primary and secondary constraints:
\begin{equation}\label{O-inv}
    \delta_\epsilon O(q,p)\approx 0 \quad \Leftrightarrow \quad
    \{O,T_\alpha\}\approx 0\,, \,\,\, \{O,\tau_a\}\approx 0\, .
\end{equation}
Also notice that the Lagrange multipliers cannot contribute to the
on-shell invariants, as $\delta_\epsilon\lambda^\alpha$ begins with
$\dot{\epsilon}{}^\alpha$ (\ref{GVar-H}), while the parameters
$\epsilon^\alpha$ are not constrained by equations (\ref{GPC-H}).

Let us now detail involution relations (\ref{GVtau}),
(\ref{dot-tau}). As $H_T=H(q,p)+\lambda^\alpha T_\alpha (q,p)$, the
structure functions $W (q,p,\lambda), \ \Gamma(q,p,\lambda)$ in
(\ref{GVtau}), (\ref{dot-tau}) are at most linear in
$\lambda^\alpha$:
\begin{equation}
\label{Walpha} W_\alpha^\beta(q,p,\lambda)
=V_\alpha^\beta(q,p)+U_{\alpha\gamma}^\beta(q,p)\lambda^\gamma\,,
\quad \Gamma_\alpha^a(q,p,\lambda) = V^a_\alpha(q,p) +
U_{\alpha\gamma}^a(q,p)\lambda^\gamma\,;
\end{equation}
\begin{equation}
\label{Wa}
W^\alpha_a(q,p,\lambda)=V_a^\alpha(q,p)-U_{\gamma
a}^\alpha(q,p) \lambda^\gamma\,, \quad \Gamma^b_a(q,p,\lambda)
=V_a^b(q,p)- U_{\gamma a}^b(q,p)\lambda^\gamma\,.
\end{equation}
By introducing uniform notation for primary and secondary
constraints $\mathcal{T}_A=(T_\alpha,\tau_a)$, $A=(\alpha, a)$, and
accounting for (\ref{Walpha}), (\ref{Wa}), the involution relations
(\ref{GVtau}), (\ref{dot-tau}) are brought to the following form:
\begin{equation}\label{InvT}
    \{\mathcal{T}_A(q,p), H(q,p)\}= V_A^B(q,p)\mathcal{T}_A(q,p)\, , \quad \{\mathcal{T}_A(q,p), \mathcal{T}_B(q,p)
    \}=U^C_{AB}(q,p)\mathcal{T}_C(q,p)\,.
\end{equation}
The above involution relations include both primary and secondary
constraints on an equal footing and merely correspond to a general
first-class system. These relations, per se, do not reveal any
indication of the equations imposed on the gauge parameters
(\ref{GPC-H}). At the level of action (\ref{SH}), however, the
differences exist as the primary constraints are included into the
action with the Lagrange multipliers, while the secondary ones are
not. It is the asymmetry which leads to equations on gauge
parameters (\ref{GPC-H}). With this regard, we mention the
long-known idea that the secondary first-class constraints $\tau_a$
can be included into the action with their own Lagrange multipliers
$\lambda^a$,
\begin{equation}\label{SH-Tscript}
   S[q,p,\lambda]= \int dt\left(p_i\dot{q}{}^i - H_{\mathcal{T}}\right)\, , \quad
   H_{\mathcal{T}}=H(q,p)+ \lambda^A\mathcal{T}_A(q,p)\,,
\end{equation}
where $\lambda^A=(\lambda^\alpha, \lambda^a)$. If we begin with this
action, it will have the usual first-order gauge symmetry,
\begin{equation}\label{GS1}
    \delta_\epsilon O(q,p)=\{ O(q,p),
    \mathcal{T}_A(q,p)\}\,\epsilon^A\,
    ,\quad\delta_\epsilon\lambda^A= \dot{\epsilon}{}^A +
    (V^A_B-U^A_{CB}\lambda^C)\,\epsilon^B \,,
\end{equation}
with unconstrained gauge parameters $\epsilon^A=(\epsilon^\alpha,\
\epsilon^a)$. The introduced multipliers $\lambda^a$ can be
considered as ``compensatory fields" to the constraints on gauge
parameters (\ref{GPC-H}) in the theory with original action
(\ref{SH}). The gauge invariants (\ref{O-inv}) of the unfree gauge
symmetry (\ref{GVar-H}), (\ref{GPC-H}) obviously coincide with the
invariants of the transformations (\ref{GS1}). At the level of
action, however, there may be a subtle difference between the theory
(\ref{SH}) with unfree gauge symmetry and the corresponding theory
with unconstrained gauge symmetry and compensatory fields
(\ref{SH-Tscript}). The matter is that the modular parameters
$\Lambda_a$ (\ref{tau-Lambda}) do not contribute to the gauge
transformations nor they are explicitly involved in the original
action (\ref{SH}). Action (\ref{SH-Tscript}) involves compensatory
fields $\lambda_a$ and secondary constrains $\tau_a$, while the
latter explicitly include modular parameters $\Lambda_a$
(\ref{tau-Lambda}). So, the action (\ref{SH-Tscript}) describes the
dynamics with fixed values of modular parameters, while the original
action encompasses the dynamics with entire moduli space
(\ref{KerG}), (\ref{tau-Lambda}). In the case of gravity, for
example, it would be the difference between the action of unimodular
gravity which encompasses dynamics with any value of cosmological
constant and the Einstein's action with fixed $\Lambda$. The role of
compensatory field is played in this case by lapse function, or
equivalently by $\det g$.

\section{Hamiltonian BFV-BRST formalism}

\noindent
In the BFV (Batalin-Fradkin-Vilkovisky) theory, the gauge
invariants are represented by zero ghost number BRST cohomology
classes in the so-called minimal sector of the ghost extended phase
space. For the basics of the formalism, we refer to the textbook
\cite{teitelboim1992quantization}. As we have seen in section 3, the gauge
invariants of unfree gauge symmetry (\ref{GVar-H}), (\ref{GPC-H})
should Poisson-commute on-shell to all the constraints
(\ref{O-inv}). The involution relations of primary and secondary
constraints define a general first-class constraint algebra
(\ref{InvT}), which does not reveal any specifics related to the
equations  imposed on gauge parameters (\ref{GPC-H}). This means
that in the minimal sector, the Hamiltonian BRST formalism is
constructed along the usual lines of the BFV method, while the
specifics of the unfree gauge symmetry is accounted for by the
non-minimal sector.

To begin with the Hamiltonian BRST embedding of the theory, we
briefly describe the minimal ghost sector in the BFV formalism.
Every first-class constraint $\mathcal{T}_A$ be it primary, or
secondary, is assigned with a pair of canonically conjugate
ghosts\footnote{For simplicity, we assume the original variables to
be even, so the ghosts are Grassmann odd.} with usual ghost numbers
\begin{equation}\label{ghosts}
\text{gh}\,C^A=-\text{gh}\,\bar{P}_A=1\, ,\quad \{C^A,P_B\}=\delta^A_B\,.
\end{equation}
The BRST charge in the minimal sector is defined as
\begin{equation}\label{Q-min}
    Q_{\texttt{min}}(q,p,C,\bar{P})=C^A \mathcal{T}_A \, +\,
    \ldots\,,\quad  \text{gh}\,Q_{\texttt{min}}=1\,, \quad \{Q_{\texttt{min}},Q_{\texttt{min}}\}=0 \,
    ,
\end{equation}
where $\ldots$ mean $\bar{P}$-depending terms that are iteratively
defined by the equation $\{Q_{\texttt{min}},Q_{\texttt{min}}\}=0$.
Any gauge invariant $O(q,p)$ (\ref{O-inv}), including
 $H$, is extended by ghosts  to become BRST-invariant:
\begin{equation}\label{H}
    H(q,p)\,\mapsto\, \mathcal{H}(q,p,C,\bar{P})=H\,+\, \ldots\,, \quad \text{gh}\,\mathcal{H}=0\,,\quad
    \{Q_{\texttt{min}},\mathcal{H}\}=0 \, .
\end{equation}
Let us discuss the non-minimal sector that is needed for the gauge
fixing. The original action (\ref{SH}) and unfree gauge
transformations (\ref{GVar-H}), (\ref{GPC-H}) involve the Lagrange
multipliers to the primary constraints only. The number of
independent gauge parameters (if they could be extracted by
resolving the equations (\ref{GPC-H}) as explained in section 3)
should be equal to the number of primary constraints. Hence, the
same number of independent conditions should be imposed for gauge
fixing. Therefore, the non-minimal sector of the theory includes the
Lagrange multipliers $\lambda^\alpha$ to the primary constraints,
and the Lagrange multipliers $\pi_\alpha$ to the independent
relativistic gauge conditions
$\dot\lambda{}^\alpha-\chi^\alpha(q,p)=0$. The corresponding
canonical ghost pair is introduced for every pair of the Lagrange
multipliers, so the complete non-minimal sector reads
\begin{equation}\label{lambda-pi}
\text{gh}\,\lambda^\alpha=\text{gh}\,{\pi}_\alpha=0\,, \quad
\text{gh}\,P_\alpha=-\text{gh}\,\bar{C}_\alpha=1\, , \quad
\{\lambda^\alpha ,
\pi_\beta\}=\{P^\alpha,\bar{C}_\beta\}=\delta^\alpha_\beta \, .
\end{equation}
Given the extended set of variables, the complete BRST charge reads
\begin{equation}\label{Q-ext}
Q= Q_{\texttt{min}}+ \pi_\alpha P^\alpha \, .
\end{equation}
With this charge, the gauge-fixed BRST invariant Hamiltonian is
defined in the usual way,
\begin{equation}\label{H-Psi}
    H_{\Psi}= \mathcal{H}+ \{Q,\Psi\}\,,\quad \Psi=\bar{C}_\alpha\chi^\alpha+ \lambda^\alpha\bar{P}_\alpha\, .
\end{equation}
The partition function $Z_\Psi$, being defined by the Hamiltonian
$H_\Psi$, does not depend on the choice of gauge conditions
included in $\Psi$ due to usual reasons of the Hamiltonian BRST
formalism \cite{teitelboim1992quantization}.

Consider $Z_\Psi$ for the simplest case when the Hamiltonian
$H_\Psi$ is at most squared in ghost variables. The path integral
for the partition function reads:
\begin{equation}\nonumber
Z_\Psi=\int \big[\mathcal{D}\varphi\big]
\exp\Big\{\frac{i}{\hbar}\int dt\,\Big(
p\dot{q}  - H(q,p) - \lambda^\alpha T_\alpha +\pi_\alpha(\dot{\lambda}^\alpha - \chi^\alpha)
+\bar{P}_a \big( \dot{C}^a +  \Gamma_b^a C^b +  \Gamma_\alpha^a C^\alpha\big)
\end{equation}
\begin{equation}\label{Zpsi}
 - \bar{C}_\alpha \big( \{ \chi^\alpha , T_\beta\} C^\beta +  \{\chi^\alpha , \tau_a  \}  C^a \big)
+\bar{P}_\alpha \big( \dot{C}^\alpha + W_\beta^\alpha C^\beta + W_a^\alpha C^a \big)
  + P^\alpha ( \bar{P}_\alpha + \dot{\bar{C}}_\alpha)
\Big)\Big\}\,,
\end{equation}
where $\varphi=\big\{q, p, \lambda^\alpha, \pi_\alpha, C^a, \bar{P}_a,
C^\alpha, \bar{P}_\alpha, P^\alpha, \bar{C}_\alpha\big\}$.
The integral by $P^\alpha$ results in $\delta( \bar{P}_\alpha +
\dot{\bar{C}}_\alpha)$, which removes the integral over
$\bar{P}_\alpha $. The result reads
\begin{equation}\nonumber
Z_\Psi=\int \big[\mathcal{D}\varphi'\big]\exp\Big\{\frac{i}{\hbar}\int dt\, \Big(
p\dot{q}  - H(q,p) - \lambda^\alpha T_\alpha +\pi_\alpha(\dot{\lambda}^\alpha - \chi^\alpha) +\bar{P}_a \big( \dot{C}^a +  \Gamma_b^a C^b +  \Gamma_\alpha^a C^\alpha  \big)
\end{equation}
\begin{equation}
 - \bar{C}_\alpha \big( \{ \chi^\alpha , T_\beta  \} C^\beta+  \{\chi^\alpha , \tau_a  \}  C^a \big)
-\dot{\bar{C}}_\alpha\big( \dot{C}^\alpha + W_\beta^\alpha C^\beta + W_a^\alpha   C^a \big)
\Big)\Big\}\,,
\end{equation}
where $\varphi'=\big\{q, p, \lambda^\alpha, \pi_\alpha, C^a, \bar{P}_a, C^\alpha,
\bar{C}_\alpha\big\}$.
The integral over anti-ghosts $\bar{P}_a$ would enforce constraints
$\dot{C}^a +  \Gamma_b^a C^b +  \Gamma_\alpha^a C^\alpha=0$. This is quite a natural
phenomenon: once gauge variations (\ref{GVar-H}) induced by
primary and secondary constraints are unfree, being restricted by equations (\ref{GPC-H}), the ghosts should obey the same
conditions as the gauge parameters do. The constraint on ghosts is the
cornerstone for the extension of the BV formalism for the theories
with unfree gauge symmetry \cite{KAPARULIN2019114735},
\cite{Kaparulin2019}. Here, we see that they naturally arise from
the Hamiltonian BFV-BRST quantization.

\section{Example: linearized unimodular gravity}

\noindent
Consider the action of unimodular
gravity linearized in the vicinity of Minkowski space background
\begin{equation}\label{S-LUG}
S \ = \ \frac{1}{4}\int d^4x \Big(\partial_{\bar{\mu}}h_{\bar{\nu}\bar{\rho}}\partial^{\bar{\mu}} h^{\bar{\nu}\bar{\rho}}  -  2\, \partial_{\bar{\mu}}  h_{\bar{\nu}\bar{\rho}} \partial^{\bar{\nu}}h^{\bar{\mu}\bar{\rho}}\Big)\,,
\quad
\eta^{\bar{\alpha}\bar{\beta}} h_{\bar{\alpha}\bar{\beta}}=0\,,
\end{equation}
where $\bar{\alpha}=0,1,2,3$,
$\eta_{\bar{\alpha}\bar{\beta}}=diag(1,-1,-1,-1)$. Gauge identity
(\ref{GI}) for (\ref{S-LUG}) reads:
\begin{equation}\label{GI-LUG}
2\,\partial_{\bar{\alpha}}\frac{\delta\, S}{\delta\, h_{\bar{\alpha}\bar{\beta}}}-\partial^{\bar{\beta}}\tau \equiv 0 \,,
\qquad
\tau=\frac{1}{2}\,\big(\partial_{\bar{\mu}}\partial_{\bar{\nu}} h^{\bar{\mu}\bar{\nu}}\big)\,,
\end{equation}
cf. (\ref{GI-eg}). Once $\partial^{\bar\beta}\tau\approx 0 $, $\tau$
is a constant on-shell, so we have $\tau -\Lambda \approx 0$, where
specific value of the constant $\Lambda$ is determined by the
asymptotics of $h$, not by Cauchy data. If the boundary conditions
admit the growing solutions, then $\Lambda$ can be non-vanishing. In
particular, there is a solution,
\begin{equation}\label{solL}
h_{\bar{\alpha}\bar{\beta}}=\stackrel{(0)}{h}_{\bar{\alpha}\bar{\beta}}
+\, \Lambda\Big(x_{\bar\alpha}x_{\bar\beta}-\frac{\eta_{\bar{\alpha}\bar{\beta}}}{4}x^2\Big)\,,
\end{equation}
with $\tau(h)=\Lambda\neq 0$, where $\stackrel{(0)}{h}$ is any
solution vanishing at infinity. Minkowski space solutions (\ref{solL}) approximate, in a sense,  the solutions of unimodular gravity
with (Anti-)de Sitter asymptotics. The higher spin analogues
\cite{SKVORTSOV2008301}, \cite{Campoleoni2013} admit similar
solutions. For higher spins, this may be even more essential because
the cosmological constant plays the role of interaction parameter
for $s>2$.

Given the gauge identity (\ref{GI-LUG}), which involves the
completion function $\tau$,  the action (\ref{S-LUG}) should enjoy
unfree gauge symmetry. It does, in full accordance with the general
prescription (\ref{GVar}), (\ref{GPC}):
\begin{equation}\label{Sinv-LUG}
\delta_\epsilon
h_{\bar{\alpha}\bar{\beta}}=\partial_{\bar{\alpha}}\epsilon_{\bar{\beta}}+
\partial_{\bar{\beta}}\epsilon_{\bar{\alpha}}-\frac{1}{2}\eta_{\bar{\alpha}\bar{\beta}}\partial_{\bar{\gamma}}\epsilon^{\bar{\gamma}}\,, \quad \delta_\epsilon S \equiv \int d^4x\, \partial_{\bar\alpha}\epsilon^{\bar\alpha}\tau \,,
\end{equation}
cf. (\ref{GS-eq}). So, the action is gauge invariant off-shell under
the condition $\partial_{\bar\alpha}\epsilon^{\bar\alpha}=0$.

By Legendre transform of (\ref{S-LUG}), we get the Hamiltonian action
\begin{equation}\label{SH-Lug}
S[h,\Pi,\lambda]=\int \ d^4x \Big(
\Pi^{\alpha\beta}\dot{h}_{\alpha\beta}- H(h,\Pi) -\lambda^\alpha
T_\alpha(\Pi)\Big),
\end{equation}
\begin{equation}\label{HT-lug}
H=  \Pi_{\alpha\beta}  \Pi^{\alpha\beta} -
\frac{1}{2}\,\Pi^2 + \frac{1}{4}\Big(2\, \partial_\alpha h_{\beta \gamma}
\partial^\beta h^{\alpha \gamma} - \partial_\alpha {h}
\,\partial^\alpha{h} -\partial_\alpha h_{\beta\gamma}
\partial^\alpha h^{\beta\gamma} \Big),
\quad T_\alpha= - 2\,\partial_\gamma \Pi^\gamma{}_\alpha\, ,
\end{equation}
where  $\alpha,\beta=1,2,3,\ \eta_{\alpha\beta} =
-\delta_{\alpha\beta}, \ h = \eta^{\alpha\beta}h_{\alpha\beta}, \
\Pi = \eta_{\alpha\beta}\Pi^{\alpha\beta}$,
$\lambda^\alpha=h^{0\alpha}$. Conservation of primary constraints
$T_\alpha$ leads to the secondary constraint
\begin{equation}\label{tau-lug}
\dot{T}_\alpha =  \{T_\alpha, H\} = -\,\partial_\alpha \tau_0 =0\,,
\quad \tau_0=\partial_\beta\partial_\gamma h^{\beta\gamma}
-\partial_\gamma \partial^\gamma {h}\,.
\end{equation}
Once $\partial_\alpha\tau_0\approx 0$, hence $\tau_0-\Lambda_0\approx 0$,
where the constant $\Lambda_0$ is determined by asymptotics of $h$ at
infinity. Secondary constraint $\tau_0$ in (\ref{tau-lug}) will
coincide with completion function $\tau$ in Lagrangian formalism
(\ref{GI-LUG}) if the the second time derivatives are excluded from
$\partial_{\bar{\mu}}\partial_{\bar{\nu}} h^{\bar{\mu}\bar{\nu}} $
by using Lagrangian equations. Involution relations (\ref{tau-lug})
correspond to the spacial components of gauge identity (\ref{GI-LUG}).
The secondary constraint conserves by virtue of the
primary ones:
\begin{equation}\label{tau-H-lug}
\dot{\tau}_0 =\{ \tau_0, H\} = -\,\partial^\alpha T_\alpha\,.
\end{equation}
This relation corresponds to the time component of gauge
identity (\ref{GI-LUG}). All the constraints Poisson-commute to each
other. The general involution relations (\ref{GIH-alpha}),
(\ref{GIH-a}) define unfree gauge transformations in Hamiltonian
formalism by the rule (\ref{GVar-H}), (\ref{GPC-H}). Substituting
specific constraints and structure coefficients of involution
relations of the unimodular gravity (\ref{HT-lug}), (\ref{tau-lug}),
(\ref{tau-H-lug}) into the general recipe (\ref{GVar-H}),
(\ref{GPC-H}),  we arrive at the unfree gauge symmetry of this
theory:
\begin{equation}\label{GVarH-lug}
\delta_\epsilon h_{\alpha\beta}\, =\,\partial_\alpha\epsilon_\beta +\partial_\beta\epsilon_\alpha\,,\quad \delta_\epsilon\Pi^{\alpha\beta}=\,-\partial^\alpha\partial^\beta\epsilon^0+\eta^{\alpha\beta}\partial_\gamma\partial^\gamma\epsilon^0\,,
\quad \delta_\epsilon\lambda^\alpha= \dot{\epsilon}^\alpha + \partial^\alpha \epsilon^0\,,
\end{equation}
\begin{equation}\label{Tdiff}
\dot{\epsilon}^0+\partial_\alpha\epsilon^\alpha=0 \, .
\end{equation}
This symmetry can be verified by direct computation. Variation
(\ref{GVarH-lug}) of action (\ref{SH-Lug}) reads
\begin{equation}\label{direct-var-lug}
\delta_\epsilon S \equiv \int d^4x
\left((\dot{\epsilon}^0 +\partial_\alpha \epsilon^\alpha) \tau_0 - \partial_0( T_\alpha \epsilon^\alpha + \tau_0 \epsilon^0) \right).
\end{equation}
It is a symmetry indeed once $\epsilon^0$ obeys equation
(\ref{Tdiff}). As we see, the general procedure of Section 4
identifies the linearized transverse diffeomorphism
(\ref{GVarH-lug}), (\ref{Tdiff}) as the gauge symmetry of
Hamiltonian action (\ref{SH-Lug}), (\ref{HT-lug}).

Consider the BFV construction for the model following the general
prescription of Section 4. The ghosts of minimal sector are assigned
to all the constraints (cf. (\ref{ghosts})), while the non-minimal
sector is assigned only to the primary constraints (see
(\ref{lambda-pi})). The BRST charge (\ref{Q-min}), (\ref{Q-ext})  for
the linearized unimodular gravity reads:
\begin{equation}\label{Q-lug}
Q=-2\,C^\alpha\partial_\beta\Pi^\beta{}_\alpha + C^0(\partial_\beta\partial_\gamma h^{\beta\gamma}-\partial_\gamma \partial^\gamma {h}-\Lambda_0)
+\pi_\alpha P^\alpha\,.
\end{equation}
Impose three independent gauge fixing conditions,
\begin{equation}\label{chi-lug}
\partial_{\bar\beta}h^{\bar{\beta}\alpha}\,\equiv\,\dot{\lambda}^\alpha - \chi^\alpha = 0\,, \quad \chi^\alpha =-\,\partial_\beta h^{\beta\alpha}\,.
\end{equation}
Introduce gauge fermion $\Psi=\bar{C}_\alpha\chi^\alpha+\lambda_\alpha\bar{P}^\alpha$, and define
gauge-fixed Hamiltonian $H_\Psi$ (\ref{H-Psi}),
\begin{equation}\label{HPsi-lug}
H_{\Psi}= H(h,\Pi) - C^0 \partial^\alpha\bar{P}_\alpha - C^\alpha\partial_\alpha \bar{P}_0 + \partial_\beta \bar{C}_\alpha\partial^\beta C^\alpha + \partial_\beta\bar{C}_{\alpha}\partial^{\alpha}C^\beta - \pi_\alpha\partial_\beta h^{\beta\alpha}+\lambda^\alpha T_\alpha -
{P}^\alpha \bar{P}_\alpha\,,
\end{equation}
where $H$ is the original Hamiltonian (\ref{HT-lug}). For $H_\Psi$
(\ref{HPsi-lug}), partition function (\ref{Zpsi}) reads
\begin{equation}\nonumber
Z_\Psi=\int \big[\mathcal{D}\varphi\big]
\exp\Big\{\frac{i}{\hbar}\int d^4x\big(
\Pi^{\alpha\beta}\dot{h}_{\alpha\beta} -H(h,\Pi) - \lambda^\alpha T_\alpha   +\pi_\alpha(\dot{\lambda}^\alpha  + \partial_\beta h^{\beta\alpha})\end{equation}
\begin{equation}
+ \bar{P}_0(\dot{C}^0 +\partial_\alpha C^\alpha) +\bar{C}_\alpha(\partial_\beta\partial^\beta C^\alpha +\partial^\alpha\partial_\beta C^\beta) +
\bar{P}_\alpha(\dot{C}^\alpha + \partial^\alpha C^0) + {P}^\alpha( \bar{P}_\alpha + \dot{\bar{C}}_\alpha) \big)\Big\}\,,
\end{equation}
where $\varphi=\big\{h_{\alpha\beta}, \Pi^{\alpha\beta},
\lambda^\alpha, \pi_\alpha, C^0, \bar{P}_0, C^\alpha,
\bar{P}_\alpha, P^\alpha, \bar{C}_\alpha\big\}$; $H$ and $T_\alpha$
are the original Hamiltonian and primary constraints (\ref{HT-lug}).
Integrating over $P^\alpha,  \bar{P}_\alpha,  \Pi^{\alpha\beta}$ we
get Lagrangian representation for $Z$,
\begin{equation}\label{Z-lug}
Z_\Psi=\int
\big[\mathcal{D}\phi'\big] \exp\Big\{\frac{i}{\hbar}\int d^4x\Big(
\mathcal{L} + \pi_\alpha\partial_{\bar{\beta}} h^{\bar{\beta}\alpha}
+ \bar{P}_0\partial_{\bar{\alpha}}C^{\bar{\alpha}} +
\bar{C}_\alpha\Box C^\alpha \Big)\Big\}\,, \quad \Box=\partial_{\bar{\mu}}\partial^{\bar{\mu}}\,,\end{equation}
where $\phi'=\big\{h_{\bar{\alpha}\bar{\beta}}, \pi_\alpha,
\bar{P}_0, C^{\bar{\alpha}}, \bar{C}_\alpha\big\}$, and
$\mathcal{L}$ is the original Lagrangian. This representation of
partition function has been deduced by Hamiltonian BFV-BRST
quantization of the model. It appears to be a reasonable adjustment
of Faddeev-Popov (FP) recipe to the case. Among the ghost terms,
the first one represents constraint imposed on ghosts, with
$\bar{P}_0$ being the Lagrange multiplier. The ghost constraint
mirrors the transversality condition imposed on the diffeomorphisms
(\ref{Tdiff}). As the gauge parameters are unfree, it is natural to
have the corresponding ghosts constrained.  The FP term
(\ref{Z-lug}) is not Poincar\'e covariant because the gauge is fixed
by  independent condition (\ref{chi-lug}), being $3d$ vector. If the
gauge condition was a $4d$ vector, the vector components would have
to be redundant, to avoid ``over-rigid" gauge fixing. This would
require some extra ghosts. In the covariant formalism, this issue is
considered in \cite{Kaparulin2019}, while the Hamiltonian analogue
will be addressed elsewhere.

\section{Concluding remarks}

\noindent 
The field theories with unfree gauge symmetry represent a
special class of models where the gauge parameters have to obey
differential equations. Every known example of these theories (see
\cite{BUCHMULLER1988292}-\cite{Francia2017} and references therein)
admits an ``almost equivalent'' analogue without constraints on
gauge parameters. The subtle difference is that the models with
unfree gauge symmetry comprise dynamics with arbitrary modular
parameters, which are involved as integration constants, while the
analogues explicitly involve fixed modular parameters. The example
of such a parameter is a cosmological constant in unimodular
gravity. It is the distinction which is behind the constraints on
gauge parameters (\ref{GPC-H}). In terms of Hamiltonian formalism,
these constraints on gauge parameters have been previously unknown
even in the examples, not to mention the general theory. We have
worked out the general Hamiltonian BFV-BRST formalism with a due
account for the unfree gauge symmetry. As we see by examples, it
corresponds well to the extension of BV method to the unfree gauge
symmetry \cite{Kaparulin2019}, though these two schemes do not
mirror each other. In the BV scheme the equations on parameters are
directly accounted for as constraints on the corresponding ghosts,
while the Hamiltonian formalism accounts for the conditions
(\ref{GPC-H}) indirectly, by an adjusted structure of the
non-minimal ghost sector.

\noindent \textbf{Acknowledgements}. We thank D.~Kaparulin and
A.~Sharapov for discussions. The work benefited from Tomsk State
University Competitiveness Improvement Program. The work of SLL is
supported by the project 3.5204.2017/6.7 of Russian Ministry of
Science and Education.

\end{document}